\documentclass[twocolumn,floatfix,prl,showpacs,superscriptaddress]{revtex4}
\usepackage{graphicx}
\usepackage{times,xspace}
\usepackage{amsbsy,amssymb,amsmath,bm}
\usepackage{graphicx,color,epsfig,rotate}
\usepackage{fancyhdr}

\def\bbbc{{\mathchoice {\setbox0=\hbox{$\displaystyle\rm C$}\hbox{\hbox
to0pt{\kern0.4\wd0\vrule height0.9\ht0\hss}\box0}}
{\setbox0=\hbox{$\textstyle\rm C$}\hbox{\hbox
to0pt{\kern0.4\wd0\vrule height0.9\ht0\hss}\box0}}
{\setbox0=\hbox{$\scriptstyle\rm C$}\hbox{\hbox
to0pt{\kern0.4\wd0\vrule height0.9\ht0\hss}\box0}}
{\setbox0=\hbox{$\scriptscriptstyle\rm C$}\hbox{\hbox
to0pt{\kern0.4\wd0\vrule height0.9\ht0\hss}\box0}}}}

\renewcommand{\vec}[1]{{\bm{#1}}}

\begin{document}
\title{Spin Dynamics of NiCl$_2$-4SC(NH$_2$)$_2$ in the Field-Induced
Ordered Phase }
\author{S. A. Zvyagin}
\affiliation{Dresden High Magnetic Field Laboratory (HLD),
Forschungszentrum Dresden - Rossendorf, 01314 Dresden, Germany}
\author{J. Wosnitza}
\affiliation{Dresden High Magnetic Field Laboratory (HLD),
Forschungszentrum Dresden - Rossendorf, 01314 Dresden, Germany}
\author{A. K. Kolezhuk}
\thanks{On leave from: Institute of
Magnetism, National Academy of Sciences and Ministry of Education,
03142 Kiev, Ukraine.}
\affiliation{Institut f\"{u}r Theoretische
Physik C, RWTH Aachen, 52056 Aachen, Germany}
\author{V. S. Zapf}
\affiliation{National High Magnetic Field Laboratory, Los Alamos
National Laboratory, MS-E536, Los Alamos, NM 87545, USA}
\author{M. Jaime}
\affiliation{National High Magnetic Field Laboratory, Los Alamos
National Laboratory, MS-E536, Los Alamos, NM 87545, USA}
\author{A. Paduan-Filho}
\affiliation{Instituto de Fisica, Universidade de Sao Paulo,
05315-970 Sao Paulo, Brazil}
\author{V. N. Glazkov}
\affiliation{P.L. Kapitza Institute for Physical Problems RAS,
117334 Moscow, Russia}
\author{S. S. Sosin}
\affiliation{P.L. Kapitza Institute for Physical Problems RAS,
117334 Moscow, Russia}
\author{A. I.~Smirnov} \affiliation{P.L.
Kapitza Institute for Physical Problems RAS, 117334 Moscow,
Russia}

\begin{abstract}
$\rm NiCl_2$-$\rm 4SC(NH_2)_2$ (known as DTN) is a spin-$1$
material with a strong single-ion anisotropy that is regarded as a
new candidate  for Bose-Einstein condensation (BEC)
of spin degrees of freedom. We present a systematic study of the low-energy
excitation spectrum of DTN  in the field-induced magnetically
ordered phase by means of high-field electron spin resonance
 measurements at temperatures down to 0.45 K. We argue that two gapped modes
observed in the experiment can be consistently interpreted within
a four-sublattice antiferromagnet model with a finite
interaction between two tetragonal subsystems and unbroken axial
symmetry. The latter is crucial  for the interpretation of the
field-induced ordering in DTN in terms of BEC. 
\end{abstract}

\maketitle

\paragraph{Introduction.--}
Field-induced phase transitions in  magnets have recently
received a considerable amount of attention
\cite{Batyev,Affleck,Nikuni-DTN,Radu,Jaime-DTN,Paduan-DTN,Zapf-DTN},
particularly in the context of the so-called Bose-Einstein
condensation (BEC) of spin degrees of freedom. In accordance to the BEC scenario, for a uniform
gas of identical particles, at some finite temperature $T_c$ when the de Broglie wavelength becomes
comparable to the average distance between the particles, a
macroscopic fraction of the gas can be ``condensed" into a single
coherent quantum state and BEC occurs. It is worth mentioning that although there are some
important arguments \cite{Coldea2-DTN} about the
interpretation of the field-induced transitions in magnets in
terms of BEC  (based on its original, canonical definition
\cite{Einstein-DTN}),  the application of this formalism
to grand-canonical ensembles of quasi-particles in  magnets
appears to be widely accepted.   An important property of BEC is the
presence of $\rm U(1)$ symmetry, which corresponds to the global
rotational symmetry of the bosonic field phase. Below $T_c$, the
$\rm U(1)$ symmetry spontaneously gets broken,  the wave function of the condensate gets coherent on a
macroscopic scale and, as a consequence, a gapless Goldstone mode
is acquired. The  model of BEC in magnets assumes that
the spin-Hamiltonian is axially symmetric with respect to the
magnetic field, requiring  $\rm U(1)$
rotational symmetry above $T_c$.

In accordance with mean-field BEC theory, the phase-diagram
boundary for a three-dimensional magnet should obey a power-law
dependence, $H - H_{c1} \sim T^{3/2}_c$, where $H_{c1}$ is the first critical field as $T\rightarrow 0$.
Importantly,  critical exponents  extracted from phase diagrams  alone can not be regarded as sufficient criteria
for identifying field-induced transitions as BEC.
For instance, the re-opening of the energy gap in the excitation spectrum of $\rm TlCuCl_3$ (which, based on analysis of critical exponents, was regarded as the best  realization of BEC of spin degrees of freedom in  magnets \cite{Nikuni-DTN}) in the
field-induced ordered state \cite{TlCuCl3-DTN} is a clear evidence for a broken uniaxial symmetry,
 which  rules out the description of the magnetic ordering in this compound in terms of BEC.

$\rm NiCl_2$-$\rm 4SC(NH_2)_2$ (known as DTN) is a gapped $S=1$
system with a single-ion anisotropy $D$ dominating over the
exchange coupling $J$ \cite{Paduan-DTN,Zapf-DTN}, and is
 a new candidate  for BEC of spin degrees of freedom.
Although below $T_c\leq 1.2$~K DTN exhibits a field-induced antiferromagnetic (AF) ordering, with critical fields
$B_{c1}=2.1$~T and $B_{c2}=12.6$~T  and
exponents corresponding to the BEC
scenario \cite{Zapf-DTN},  the low-energy excitation spectrum
in DTN (and, correspondingly, the microscopical picture
of magnetic interactions) in the field-induced ordered state
still remains an open question. In the present work,
we have addressed these issues experimentally  by means of electron spin resonance (ESR) measurements performed at 
temperatures down to 0.45 K. Two gapped modes were
observed.  Based on  our detailed analysis, it is argued that
 the above excitation spectrum can be consistently interpreted within a four-sublattice
AF model with an intact axial symmetry (at least on the energy scale down to 1.2 K, which corresponds to the lowest frequency used in our experiments, 25 GHz). The latter is of
particular importance, being a necessary prerequisite for the
interpreting of the AF ordering in DTN in terms of the BEC
scenario.

DTN is characterized by the $I4$ space group \cite{Paduan2-DTN}
with a body-centered tetragonal lattice that may be viewed as two interpenetrating tetragonal
subsystems (hereafter TS). At $B \parallel c$ the spin dynamics can be described by the spin-Hamiltonian
\begin{equation}
\label{Ham1}
\mathcal{H}_{0}=\frac{1}{2}\sum_{\vec{n}, \vec{\delta}}
J_{\vec{\delta}}\vec{S}_{\vec{n}}\cdot
\vec{S}_{\vec{n}+\vec{\delta}} +D \sum_{\vec{n}}(S^z_{\vec{n}})^2
- h \sum_{\vec{n}} S_{\vec{n}}^z +\mathcal{H}_{\rm int},
\end{equation}
where $\vec{S}_{\vec{n}}$ are spin-1 operators at site $\vec{n}$, the
vectors $\vec{\delta}$ connect the site $\vec{n}$ to its nearest neighbors
within the same subsystem,  $h=g_c\mu_{B}B$ is the
Zeeman term, and $\mathcal{H}_{\rm int}$ describes (yet unspecified)
additional interactions. Assuming that the latter are much weaker than the
interchain interaction within TS, the Hamiltonian
parameters were estimated as $D=8.9$~K, $J_c=2.2$~K, $J_{a,b}=0.18$~K, and
$g_c=2.26$ \cite{Zvyagin2-DTN}.

\paragraph{Experimental results.--} The ESR measurements were done
at the Kapitza Institute using a transmission-type ESR
spectrometer equipped with a cylindrical multimode resonator and a $^3$He-cryostat.
High-quality single-crystalline DTN samples from the same batch as
in Ref.\ \cite{Zvyagin2-DTN} were used. The magnetic field was
applied along the tetragonal $c$ axis.

\begin{figure}
\begin{center}
\vspace{0.3cm}
\includegraphics[width=0.9\columnwidth]{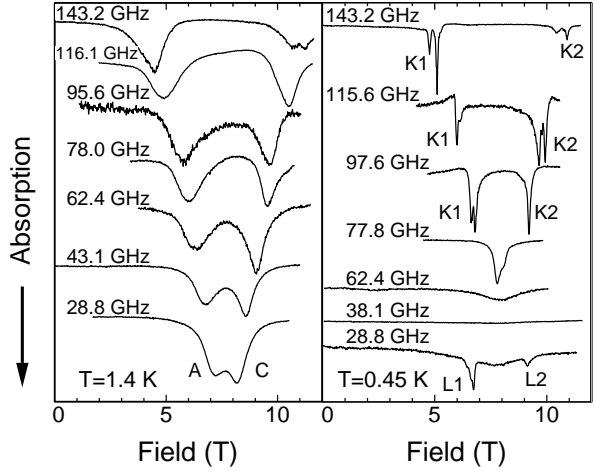}
\caption{\label{Spectra} Typical ESR spectra of DTN in the
quantum-disordered phase taken at $T=1.4$~K (left) and in the
ordered phase taken at $T=0.45$~K (right). The spectra are offset
and rescaled for clarity.}
\end{center}
\end{figure}

\begin{figure}
\begin{center}
\includegraphics[width=0.9\columnwidth]{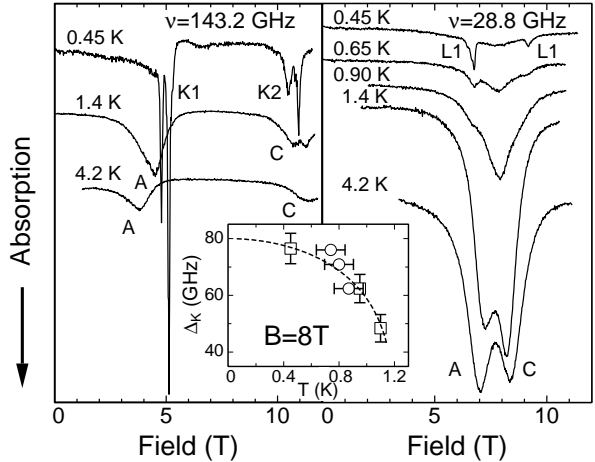}
\caption{\label{temper} Temperature dependencies of ESR spectra
taken at $\nu=143.2$~GHz (left) and 28.8~GHz (right); the
absorption scale is retained at each frequency. The inset shows
the temperature evolution of the energy gap $\Delta_K$ of the $K$
mode, at magnetic field of $8$~T (the dashed line is a guide to
the eye).}
\end{center}
\end{figure}

\begin{figure}
\begin{center}
\vspace{-0.3cm}
\includegraphics[width=0.5\textwidth]{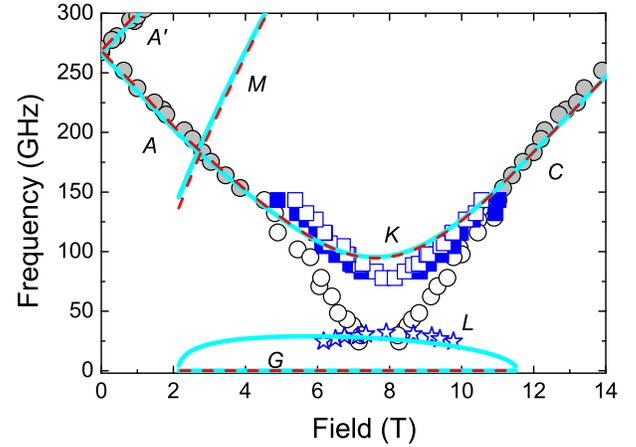}
\vspace{-1.0 cm}
\caption{\label{FFD-BEC} (Color online) The
frequency-field dependence of ESR in DTN measured at
$T=1.4$~K (open circles) and $0.45$~K (squares and stars). Filled
grey circles  denote the high frequency data ($T=1.6$~K) taken
from Ref. \cite{Zvyagin2-DTN}. The  dashed (red) lines
correspond to results of calculations for the simplest axially
symmetric two-sublattice AF model with parameters $g_c=2.26$,
$D=9.4$~K, $\widetilde{J}_{0}=2.0$~K. The  solid (cyan) lines  correspond to
results of model calculations assuming  $d=0.02$~K (see text for details).}
\end{center}
\end{figure}

The examples of the ESR spectra taken at $1.4$~K are shown in
Fig.\ \ref{Spectra} (left), while the ESR
temperature evolution is illustrated in Fig. \ref{temper}. The
corresponding frequency-field dependencies of
magnetic excitations are presented in Fig.\ \ref{FFD-BEC}. Two ESR
lines (denoted as $A$ and $C$) have been observed  at $1.4$~K. Comparison with the data of Ref.\
\cite{Zvyagin2-DTN} (filled grey circles, Fig.\ \ref{FFD-BEC})
reveals that the mode $A$ continues smoothly from the ESR line in
the low-field disordered phase,  while the mode $C$ corresponds to single-magnon
excitations in the high-field phase.

Several important features have been observed at lower
temperatures. Upon cooling down, the modes $A$ and $C$  shift towards each other (Fig.\ \ref{temper}). While
at a temperature of $1.4$~K they seem to cross at zero frequency
in the vicinity of $B\sim 8$~T, in the low-temperature AF ordered
phase they are converted into a new gapped mode $K$. This mode exhibits a slight but distinct splitting, the origin of which will be
discussed below  (the corresponding frequency-field dependence of
the resonances is denoted in Fig.\ \ref{FFD-BEC} by pairs of open
and closed squares).  The
temperature dependence of the gap $\Delta_K$ at $B=8$~T has been obtained
by measuring the temperature of the maximum absorption at a
fixed field (i.e., by the observation
of so-called ``temperature resonance") or by using a
conventional ESR procedure, recording transmission \emph{vs}
magnetic field  at different temperatures (circles and squares in
Fig.\ \ref{temper}, inset, correspondingly). The diminishing of
$\Delta_K$ upon warming can be explained by the decrease of the
order parameter. The low temperature spectrum demonstrates one more set of
resonances appearing at low frequencies (mode $L$, denoted by
stars in Fig.\ \ref{FFD-BEC}). Unlike
mode $K$, the integrated intensity of line $L$ is  $50 -
100$ times smaller than that of modes outside
the AF phase, e.g. modes $A$ and $C$.

\paragraph{Discussion.--}
The observation of the gapped mode $L$ and splitting of mode $K$
in a field range of $6\div 10$~T clearly indicates the presence of
additional interactions, that are not accounted for in the
simplest axially symmetric two-sublattice AF model (which would
have the Goldstone mode as its lowest energy excitation and
a  single mode $K$ as the next excitation branch as
indicated by the dashed (red) lines in Fig.\ \ref{FFD-BEC}).  To
 gain further insight, we have studied several
possible mechanisms that might cause gaps
in the ESR excitation spectrum.

To do so, we have considered the model (\ref{Ham1})
with various additional interactions $\mathcal{H}_{\rm int}$, using the
mean-field approach outlined below. Assuming the wave function to be a
product of single-spin coherent states parameterized by a pair of
two-component real vectors $\vec{u}$, $\vec{v}$:
\[
|\psi\rangle = s|t_{0}\rangle
+\sum_{j=x,y}(u_{j}+iv_{j})|t_{j}\rangle,\quad s=(1-u^{2}-v^{2})^{1/2},
\]
where the spin-1 states are
$|0\rangle=|t_{0}\rangle$ and $|\pm\rangle =\mp
\frac{1}{\sqrt{2}}(|t_{x}\rangle \pm i |t_{y}\rangle)$, one
obtains the effective Lagrangian in the form
$\mathcal{L}=\hbar\sum_{\vec{n}}(
\vec{u}_{\vec{n}}\cdot\partial_{t}\vec{v}_{\vec{n}}
-\vec{v}_{\vec{n}}\cdot\partial_{t}\vec{u}_{\vec{n}}) -\langle
\mathcal{H} \rangle$, with
\begin{eqnarray}
\label{H-mf}
\langle \mathcal{H} \rangle &=&
\textstyle\sum_{\vec{n}}\big\{D(u_{\vec{n}}^{2} +v_{\vec{n}}^{2})
-2\vec{h}\cdot(\vec{u}_{\vec{n}}\times \vec{v}_{\vec{n}})\big\}
\nonumber\\
& +&
2\textstyle\sum_{\langle \vec{n}\vec{n}'\rangle}
J_{\vec{n}-\vec{n}'} \big\{
(\vec{v}_{\vec{n}}\cdot \vec{v}_{\vec{n}'})
 (s_{\vec{n}}s_{\vec{n}'}+\vec{u}_{\vec{n}}\cdot \vec{u}_{\vec{n}'})
\nonumber\\
&-&(\vec{u}_{\vec{n}}\cdot
\vec{v}_{\vec{n}'})(\vec{v}_{\vec{n}}\cdot \vec{u}_{\vec{n}'})\big\} +\langle \mathcal{H}_{\rm int}\rangle ,
\end{eqnarray}
where $\langle \vec{n}\vec{n}'\rangle$ denotes summation over
nearest neighbors within the corresponding TS.
For weak fields,  the static mean-field solution  is
$\vec{u},\vec{v}=0$. However, at  $B>B_{c1}$ one
obtains a nontrivial solution with finite $\vec{u}$
and $\vec{v}$ determined by the minimization of
$\langle\mathcal{H}\rangle$, which describes the in-plane AF order
and finite magnetization along the $z$ axis. Linearizing around
the static solution and solving the equations of motion, one can
find the excitation energies $\varepsilon_{\vec{k}}$. Outside the
AF phase the ESR modes correspond to $\varepsilon_{\vec{k}=0}$,
while inside the ordered phase, due to the folding of the
Brillouin zone, one has the additional transitions at  $\vec{Q}_{B}=(\pi/a,\pi/b,\pi/c)$ with  energies $\varepsilon_{\vec{k}=\vec{Q}_{B}}$.

First we have considered the possibility of axial symmetry
breaking, caused either by  rhombic
in-plane anisotropy or by the Dzyaloshinskii-Moriya (DM)
interaction inside a TS, with the DM vector
$\vec{d}$ deviating from the anisotropy $c$ axis (note that DM is
generally allowed in DTN due to the absence of inversion symmetry
\cite{ref1}). In both cases, the axial symmetry break-down would
open a gap in the excitation spectrum,
`lifting' the Goldstone mode. Even so, this mode
corresponds to a coherent collective  excitation of the magnon
condensate below $T_c$, and thus would be expected to be
particularly strong in comparison with ESR
excitations in the disordered phase. As mentioned, the gapped mode
$L$ observed in our experiments is much less  intense
than ESR absorptions above $T_c$,
raising questions about the above scenarios. In
addition, the rhombic in-plane anisotropy would necessarily cause
splitting of the doublet at $B=0$  (roughly of the same strength as the maximum energy
of the $L$ mode), which was not observed in the experiment. It is worth mentioning that the DM term is cubic in
$\vec{u}$, $\vec{v}$, so it does not contribute to the quadratic
spectrum at $B<B_{c1}$ and thus does not produce mode splitting at
$B=0$. However, a finite in-plane component of $\vec{d}$ (e.g.,
$d_{y}$) contributes to the energy as
$W_{DM}=\frac{1}{2}\sum_{\vec{n}\vec{\delta}} \eta_{\vec{n}} d_{y}
(\vec{S}_{\vec{n}}\times \vec{S}_{\vec{n}+\vec{\delta}})_{y}$,
where $\eta_{\vec{n}}=\pm1$ for sites $\vec{n}$ belonging to the
different magnetic sublattices (inside the same TS). Close to the fully spin-polarized state, this
contribution provides an energy gain linear in the AF order
parameter, while the corresponding Zeeman energy loss is
quadratic, so AF order is favored for arbitrarily large $B$.
Consequently (and this is of particular importance for our analysis),  such a symmetry-breaking DM term would lead to the
absence of the second critical field $B_{c2}$, which is
incompatible with results of \cite{Paduan-DTN,Zapf-DTN}.

Here  we argue that a natural explanation of
all available experimental results
 is possible if  we take into account a weak isotropic corner-center interaction
of magnetic ions in the body-centered tetragonal lattice, i.e. an
interaction between TS that
preserves the axial symmetry. In case of such interactions in addition to
  conventional  (relativistic)  modes, modes
with  antiphase oscillations of two interacting AF sublattices (exchange modes) should be present.
Having much weaker coupling to the microwave field \cite{Bar}, exchange modes should be less intensive  than relativistic modes. In accordance to our observations, the gapped mode $L$ would then
be an exchange mode, coexisting with the lowest-energy relativistic mode.

The physically simplest scenario would correspond to the isotropic
``corner-center'' exchange. However, the theoretical analysis of
such a model is very difficult; if, as indicated by previous studies \cite{Zvyagin2-DTN},
the exchange interaction within each TS is AF, then
the system is highly frustrated and its mean-field ground state at
$B>B_{c1}$ is infinitely degenerate. This degeneracy may be lifted
by quantum fluctuations or by additional interactions. For that
reason, we would like to illustrate the effect of corner-center
coupling by assuming a finite DM interaction between the TS, which
already lifts the degeneracy  at the mean-field level:
\begin{equation}
\label{DMiso}
\mathcal{H}_{\rm int}= \sum_{\vec{n} \vec{n}'} \eta_{\vec{n}}\eta_{\vec{n}'}
\vec{d}\cdot (\vec{S}_{A,\vec{n}}\times \vec{S}_{B,\vec{n}'}),
\end{equation}
where the DM vector $\vec{d}$ is along the $c$ axis and $A$, $B$
denote the two TS. Such interaction favors  a
$90^{\circ}$ angle between the AF order parameters of the
subsystems. The static mean-field solution at $B>B_{c1}$ is then
\begin{eqnarray}
\label{static}
&&\left(\begin{array}{lr} \vec{u}_{A,\vec{n}} & \vec{u}_{B,\vec{n}} \\
  \vec{v}_{A,\vec{n}} & \vec{v}_{B,\vec{n}} \end{array}\right)
=\eta_{\vec{n}}\left( \begin{array}{lr} u_{0} \vec{\widehat{x}} & u_{0}
  \vec{\widehat{y}} \\
  v_{0}\vec{\widehat{y}}  &  v_{0}\vec{\widehat{x}}\end{array}\right),
\end{eqnarray}
and the
  excitation energies $\varepsilon_{\vec{k}}$ are obtained from the secular equation
  $\det(R)=0$, with $R=\left[\begin{array}{lr}  F & U\\ U^{\dag} & F
  \end{array}\right]$, where
\[
F=\left[
\begin{array}{cccc}
-D+W_{1} & 0 & i\varepsilon_{\vec{k}} & -h+W_{4}\\
0 & -D+W_{3} & H+W_{2} & i\varepsilon_{\vec{k}}\\
-i\varepsilon_{\vec{k}} & h+W_{2} & -D+W_{5} & 0 \\
-H+W_{4} & -i\varepsilon_{\vec{k}} & 0 & -D+W_{6}
\end{array}
\right]
\]
the matrix $U$ describes intersubsystem   interaction,
\[
 U=4\widetilde{d}_{k}\left[
\begin{array}{cccc}
-s_{0}^{2} & 0 & 0 & 0 \\
0 & s_{0}^{2}-2v_{0}^{2}+v_{0}^{4}/s_{0}^{2} & -u_{0}v_{0}(1-v_{0}^{2}/s_{0}^{2})
& 0 \\
0 & -u_{0}v_{0}(1-v_{0}^{2}/s_{0}^{2}) & u_{0}^{2}v_{0}^{2}/s_{0}^{2} & 0\\
0 & 0 & 0 & 0
\end{array}
\right],
\]
and the following shorthand notations have been used:
\begin{eqnarray*}
&& W_{1}=-4 s_{0}^{2}\widetilde{J}_{k} +W_{6}, \quad
W_{2}=u_{0}v_{0}(\widetilde{f}_{k}-\widetilde{g}_{k} )
\\
&&
 W_{3}=
4\widetilde{J}_{k}(2u_{0}^{2}-1)
+v_{0}^{2}(3\widetilde{f}_{k}-\widetilde{g}_{k}),\quad
W_{4}=4u_{0}v_{0}\widetilde{J}_{0}\\
&&
W_{5}=+v_{0}^{2}\widetilde{f}_{\vec{k}}
-u_{0}^{2}\widetilde{g}_{\vec{k}}, \quad W_{6}=-4v_{0}^{2}(\widetilde{J}_{0}+\widetilde{d}_{0})
\end{eqnarray*}
with $ \widetilde{J}_{\vec{k}}= \sum_{\vec{\lambda}=\vec{a},\vec{b},\vec{c}}
J_{\vec{e}}\cos(\vec{k}\cdot\vec{\lambda})$,  $\widetilde{d}_{0}=4d$, and
\begin{eqnarray*}
&&  \widetilde{d}_{\vec{k}}\equiv (\widetilde{d}_{0}/8)
\prod_{\vec{\lambda}=\vec{a},\vec{b},\vec{c}}
\big(1-\exp(i\vec{k}\cdot\vec{\lambda}) \big) \\
&&
\widetilde{f}_{\vec{k}}=4(\widetilde{J}_{\vec{k}}-\widetilde{J}_{0}-\widetilde{d}_{0}),
\quad \widetilde{g}_{\vec{k}}=(4v_{0}^{2}/s_{0}^{2})(\widetilde{J}_{\vec{k}}+\widetilde{J}_{0}+\widetilde{d}_{0}).
\end{eqnarray*}

The results of model calculations using $g_c=2.26$, $D=9.4$~K,
$\widetilde{J}_{0}= 2.0$~K, $d=0.02$~K are shown in Fig.\
\ref{FFD-BEC} by solid (cyan) lines. Due to a finite interaction between the two
TS (the term $d$), the low-energy mode is split
into a doublet with a gapped upper component and a zero frequency
lower component. One can see that the model qualitatively
describes frequency-field dependencies of all observed ESR modes
assuming the existence of  a
 gapless Goldstone mode $G$,
which can not be detected experimentally. It is important to mention that
it was not possible to fit the frequency-field
dependence of magnetic excitations in DTN in the AF phase using
the set of parameters obtained in Ref.\ \cite{Zvyagin2-DTN},
although they are very close to those used in our calculations.
This discrepancy mainly  stems from neglecting quantum
fluctuations in the mean-field-theory approach used in the present
paper.
In the above calculation
the mode $K$ (determined by $\varepsilon_{\vec{k}=0}$) comes out
doubly degenerate,  which is an artefact of the model assumption
on purely DM inter-TS interaction, making the interaction matrix
$U$ vanish at $\vec{k}=0$. Any finite symmetric exchange
interaction between the TS will be sufficient to lift this
degeneracy and thus  will explain the
observed slight splitting of the $K$ mode. The theory also
predict  the existence of  a
third ESR mode $M$ in the AF phase (Fig.\ \ref{FFD-BEC}),
which corresponds to the second magnon branch at
$\vec{k}=\vec{Q}_{B}$; this mode could not
 be observed in the present study due to the limited
frequency range.

In summary, high-field ESR studies of magnetic
excitations in the field-induced ordered phase of DTN  have been performed  at frequencies down to 25 GHz.
Two gapped modes  were observed in the ESR
spectrum. Our experimental observations can be consistently
interpreted within the four-sublattice AF model with an intact
axial symmetry, which is of crucial importance for  the
interpretation of the field-induced  ordering in DTN in terms of
 BEC of spin degrees of freedom.

\paragraph{Acknowledgments.--}
We thank  C.D. Batista,  M. Kenzelmann,  S.
Zherlitsyn, and A. A. Zvyagin for discussions. This work was
partly supported by the DFG, Grant  ZV~6/1-1 and
by the RFBR, Grant
06-02-16509. AK is supported by the DFG Heisenberg Program, Grant KO~2335/1-2. APF is grateful for support from CNPq and FAPESP
(Brazil).


\begin{thebibliography}{10}

\bibitem{Batyev} E.~G.~Batyev and L.~S.~Braginskii, Sov. Phys. JETP {\bf 60}, 781 (1984).

\bibitem{Affleck} I.~Affleck, Phys. Rev. B {\bf 41}, 6697 (1990).

\bibitem{Nikuni-DTN} T.~Nikuni \emph{et al.},
Phys. Rev. Lett. {\bf 84}, 5868 (2000).

\bibitem{Radu} T. Radu \emph{et al.},
Phys. Rev. Lett. {\bf 95}, 127202 (2005).

\bibitem{Jaime-DTN} M.~Jaime \emph{et al.}, Phys. Rev. Lett. {\bf 93},
087203 (2004).

\bibitem{Paduan-DTN} A.~Paduan-Filho \emph{et al.},
Phys. Rev. B {\bf 69}, 020405(R) (2004).

\bibitem{Zapf-DTN} V.~S.~Zapf \emph{et al.}, Phys. Rev. Lett. {\bf 96},
077204 (2006).


\bibitem{Coldea2-DTN}  D.~L.~Mills, Phys. Rev. Lett. {\bf 98}, 039701
(2007); T.~Radu \emph{et al.}, Phys. Rev. Lett. {\bf 98}, 039702 (2007).

\bibitem{Einstein-DTN} S.~N.~Bose, Z. Phys. {\bf 26}, 178 (1924);
A. Einstein, Sitzungsber. Kgl. Preuss. Akad. Wiss. {\bf 1}, 3
(1925).

\bibitem{TlCuCl3-DTN} V. N. Glazkov \emph{et al.}, Phys.
Rev. B {\bf 69}, 184410 (2004); A.~K.~Kolezhuk \emph{et al.},
Phys. Rev. B {\bf 70}, 020403(R) (2004).

\bibitem{Zapf2-DTN} V.~S.~Zapf \emph{et al.}, J. Appl. Phys. {\bf 101},  09E106 (2007).

\bibitem{Paduan2-DTN} A.~Paduan-Filho \emph{et al.}, J. Chem. Phys. {\bf
74}, 4103 (1981).

\bibitem{Zvyagin2-DTN} S.~A.~Zvyagin \emph{et al.}, Phys.  Rev. Lett.
{\bf 98}, 047205 (2007).

\bibitem{ref1} It is worth mentioning that $\Delta
M_{s}=2$ ESR transitions from the ground state to two-magnon bound
states observed in the high-field phase \cite{Zvyagin2-DTN} indicate
a small nonconservation of $S^{z}$ quantum number, which might be a signature of the broken axial symmetry in DTN. On the other hand,
one should keep in mind that even a slight (a few degrees) misorientation
of the sample with respect to the applied field  might allow such transitions.

\bibitem{Bar} V.~G.~Bar'yakhtar  \emph{et al.}, Sov. Phys. JETP {\bf 61}, 823 (1985).


\end{thebibliography}
\end{document}